\documentclass[aps,prb,twocolumn,groupedaddress]{revtex4}

\usepackage{graphicx}
\usepackage{amsfonts}
\usepackage{amssymb}

\begin{document}


\title{Conductance plateau transitions in 
quantum Hall wires with 
spatially correlated random magnetic fields}

\author{Tohru Kawarabayashi and Yoshiyuki Ono}
\affiliation{Department of Physics, Toho University,
Miyama 2-2-1, Funabashi 274-8510, Japan}

\author{Tomi Ohtsuki}
\affiliation{Department of Physics, Sophia University,
Kioi-cho 7-1, Chiyoda-ku, Tokyo 102-8554, Japan}

\author{Stefan Kettemann}
\affiliation{I. Institut f\"{u}r Theoretische Physik, Universit\"{a}t Hamburg,
20355 Hamburg, Germany}

\author{Alexander Struck}
\affiliation{Department of Physics, University of Kaiserslautern, 
D-67663 Kaiserslautern, Germany}

\author{Bernhard Kramer}
\affiliation{Jacobs University Bremen, 28757 Bremen, Germany}

\date{\today}

\begin{abstract}
Quantum transport properties  in quantum Hall wires in the presence of 
spatially correlated disordered magnetic fields are investigated numerically. 
It is found that the correlation drastically changes the 
transport properties associated with the edge state,   
in contrast to the naive expectation that the 
correlation simply reduces the effect of disorder.  
In the presence of correlation, the separation  
between the successive conductance plateau transitions becomes larger than 
the bulk Landau level separation determined by the mean value of the 
disordered magnetic fields. The transition energies coincide with the 
Landau levels in an effective magnetic field stronger than the mean value of 
the disordered magnetic field. For a long wire, 
the strength of this effective magnetic field
is of the order of the maximum value of the magnetic fields in the system.
It is shown that the effective field is determined by a part 
where the stronger magnetic field region connects 
both edges of the wire.
\end{abstract}

\pacs{72.15.Rn, 73.20.Fz, 73.43.Nq}

\maketitle

\section{Introduction}

Since the discovery of the quantum Hall effect, 
quantum transport property
of two-dimensional(2D) systems in strong magnetic fields 
has been one of the central issues of the condensed matter
physics \cite{Ando,Aoki,PG,Hatsugai,Huckestein,KOK}. 
The energy spectrum of electrons in two dimensions in a 
strong magnetic field forms equally spaced degenerate energy levels called 
the Landau levels. In the presence of boundaries, there exist 
the edge states extending 
along the boundary of the sample \cite{Halperin,Buttiker}. 
The edge state corresponds to the classical skipping orbit along the boundary 
and is known to be less influenced by impurities and defects 
of the system \cite{MF}. Since  
the edge state is associated with each Landau level, the number of 
edge states at the Fermi energy is equal to the number of 
the Landau levels below it.  
The conductance is therefore quantized to be $n(e^2/h)$ per spin when the 
Fermi energy lies between the $n$th and the $(n+1)$th Landau levels.
In the absence of impurity scattering,
the transition between quantized values of the conductance occurs when the 
Fermi energy crosses the Landau level.

\begin{figure}
\includegraphics[scale=0.65]{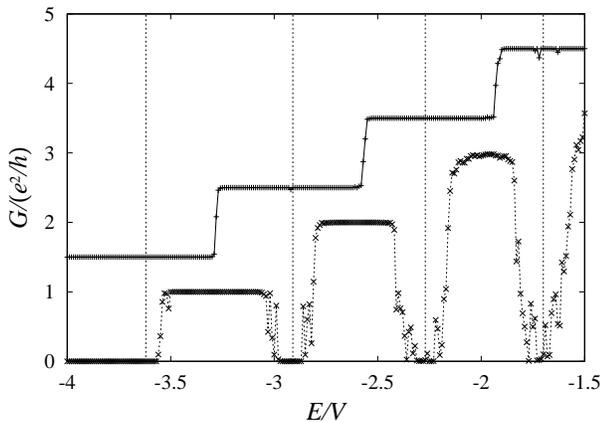}
\caption{
Conductance for the case of the potential disorder in the square lattice 
tight-binding model. The system size is 
$20 \times 1000$ and the magnetic flux $\phi$ per 
plaquette is assumed to be $\phi/(h/e) = 1/16$.
The site energies $\varepsilon_i$ obey the Gaussian distribution with 
variance $\sigma \approx 0.23$ in units of the hopping amplitude. The spatial 
correlation of site energies is assumed to be 
$\langle \varepsilon_i \varepsilon_j \rangle =\langle \varepsilon^2 \rangle
\exp (-r_{ij}^2/2\eta)$,
where $r_{ij}$ denotes the distance between the sites $i$ and $j$.
Conductance with potential correlation ($\eta/a=5$ $(+)$) and 
that without it ($\eta=0$ $(\times))$ are plotted. For $\eta/a=5$, 
$G/(e^2/h) +1.5$ is shown instead of $G/(e^2/h)$. Here the lattice 
constant is denoted by $a$. The vertical dotted lines represent the positions 
of the conductance steps in the absence of disorder, namely the 
positions of the bulk Landau levels.\cite{KOOKSK}
\label{fig1}
}
\end{figure}

The stability of the edge states and the mixing of the edge and 
the bulk states in the presence of impurities has been studied  
by many authors \cite{SKM,OOSSC,OOK,Ando2,AA}. 
It has been 
shown that even in the presence of disorder, 
the edge state is well defined and is extended 
along the boundary as long as  its energy 
lies away from those of the bulk Landau subbands and the 
magnetic field is strong enough. On the other hand, 
when its energy lies in the 
middle of the bulk Landau subbands, the edge state mixes with 
the bulk states by the impurity scattering, 
which leads to the localization of edge states 
\cite{Ando2,AA}. It has been shown for 
long quantum Hall wires with uncorrelated disorder potential 
that the conductance indeed vanishes when the Fermi energy 
is close to the centers of the bulk Landau subbands \cite{KKO,SKOK}.
The zero conductance regime therefore appears between the 
conductance plateaus $n(e^2/h)$ and $(n+1)(e^2/h)$ in the 
presence of disorder (Fig. \ref{fig1}). 
This transition between the quantized conductance 
and the insulator is called the chiral metal-insulator transition
(CMIT) \cite{KKO,SKOK}.

We focus on the effect of disorder correlation
in such systems. 
For the bulk quantum Hall system, 
the effect of potential correlation has been 
studied \cite{AndoUemura,OO2,Ando3,KS}
and demonstrated numerically that the potential correlation suppresses 
the mixing between edge states \cite{OO2}. 
It has also been 
observed that the critical energies of the bulk system are 
insensitive to the potential correlation as long as the disorder
is weak\cite{KS}.
All these results are consistent with the intuitive picture that 
the potential correlation makes the potential smooth and reduces the 
scattering with large momentum transfers.
 
It is therefore 
expected naively that in the case of the quantum Hall wires, 
the zero conductance regime observed for the 
uncorrelated potential is suppressed and  
the quantized conductance steps are recovered when
the potential correlation is introduced.
In the previous paper\cite{KOOKSK},we 
have found indeed the suppression of CMIT which yields the 
recovery of the quantized conductance steps. Quite unexpectedly, however, 
it has also been found that the quantized conductance 
steps shift toward higher energies (Fig. \ref{fig1}). This is remarkable since 
this means that the potential correlation acts against the 
conduction of the system contrary to the naive expectation.
We have found that the scale of the shift is of the order of the 
strength of random potential and is insensitive to the 
strength of the magnetic field. The shift appears in wires but does not  
appear in short systems. For understanding the physics 
behind this shift, we have argued that the correlated potential
can act as a potential barrier across the wire, which produces 
the same amount of shift for each plateau transition.
It has also been found \cite{KOOKSK2} 
by calculating the conductance for various values of 
energies keeping the disorder potential fixed, that
though the size of the shifts depends considerably 
on the sample, the distance between 
successive transitions, namely the plateau width, 
is insensitive to the sample and is essentially 
the same as the bulk Landau level separation. 
The quantum Hall wire is therefore an interesting system where the 
potential correlation affects qualitatively its transport property.

In the present paper, 
we further examine the effect of the disorder 
correlation in quantum Hall wires 
by considering a different type of disorder, namely 
the disordered magnetic fields.
When the disordered component of the magnetic fields is smaller than its 
uniform component, 
the system belongs to the quantum Hall universality class 
except for the band center\cite{Huckestein2}.  It is therefore expected that 
even in the case of the disordered magnetic field, 
the conductance plateau transitions also occur in the same way as the 
disordered potential case. 
The disordered magnetic field system has been studied theoretically 
in connection to 
the fractional quantum Hall effect at filling factor $\nu = 1/2$\cite{HLR}.
Due to the advances in the fabrication technique, such a system has also 
been realized experimentally in a semiconductor heterostructure and its 
magnetotransport has been studied under various conditions \cite{AEKI,MZMCG}.
It is therefore important to clarify the effect of correlation 
in 2D disordered magnetic fields theoretically and experimentally. 

We consider a lattice model with spatially correlated disordered 
magnetic fields and examine numerically 
how the effect of correlation shows up in the conductance 
plateau transitions in such a system.
It is then found that the effect appears as the change of the 
separation of successive plateau transitions, which is 
qualitatively different from that 
observed for the correlated disorder potential case.
The origin of this change of the separation is also discussed 
semi-classically.

\section{Models and Methods}

We consider a two-dimensional system with disordered magnetic fields
described by the Hamiltonian
on the square lattice
\begin{equation}
 H = \sum_{<i,j>} V \exp({\rm i} \theta_{i,j}) c_i^{\dagger} 
 c_j .
\end{equation}
The phases $\{ \theta_{i,j} \}$ are determined 
so that the summation around the $i$th plaquette is equal to the 
magnetic flux $-2\pi\phi_i/\phi_0$ through the plaquette, where
$\phi_0 = h/e$ stands for the flux quantum.  
The flux $\phi_i$ through the $i$th plaquette can be expressed as
\begin{equation}
 \phi_i = \delta \phi_i + \phi.
\end{equation}
Here $\phi$ denotes the uniform component of the magnetic fluxes, which 
determines the mean value of fluxes. The disordered component around the mean 
is denoted by $\delta \phi_i$. 
It is assumed that the disordered component of the flux is distributed with 
the Gaussian distribution with zero mean as
\begin{equation}
 P(\delta \phi) = \frac{1}{\sqrt{2\pi \sigma_{\phi}^2}}
 \exp(-\delta \phi^2/2\sigma_{\phi}^2). \label{Gauss}
\end{equation} 
The spatial correlation of the disordered components is 
assumed to be 
\begin{equation}
 \langle \delta \phi_i \delta \phi_j \rangle = 
 \langle \sigma_{\phi}^2 
 \rangle \exp (-|\mbox{\boldmath $R$}_i-
 \mbox{\boldmath $R$}_j|^2/4\eta_{\phi}^2). \label{corr1}
\end{equation}
The position vector for the site $i$ is denoted by $\mbox{\boldmath $R$}_i$.
All length scales are measured in units of the lattice constant.
The spatially correlated fluxes are constructed from the uncorrelated 
fluxes $\delta \phi^{\rm u.c.}_j$'s as \cite{KS,KOOKSK}
\begin{equation}
 \delta \phi_i = \frac{\sum_j \delta \phi^{\rm u.c.}_j \exp 
 (-|\mbox{\boldmath $R$}_j-
 \mbox{\boldmath $R$}_i|^2/2\eta_{\phi}^2)}{\sqrt{\sum_j
 \exp (-|\mbox{\boldmath $R$}_j-\mbox{\boldmath $R$}_i|^2/\eta_{\phi}^2)}}.
 \label{sum}
\end{equation}
When the uncorrelated fluxes $\delta \phi^{\rm u.c.}_j$'s 
obey the Gaussian distribution
with variance $\sigma$, 
it is easily verified that $\delta \phi_i$ satisfies 
the relations in eqs.(\ref{Gauss}) and (\ref{corr1}). 
The effective width $w_{\phi} = \sqrt{12}\sigma_{\phi}$ is 
used to specify the disorder strength. In the case of the present 
lattice model, 
the flux $\phi_i$ and $\phi_i + m\phi_0$ with an integer $m$ are 
equivalent with each other. 
To avoid the artifact of this periodicity \cite{Hofstadter},
we confine ourselves to small values of  
$w_\phi$ and $\phi$ and to the Fermi energy away from the band center.

We consider a system with the length $L$ and the width $M$ having two leads 
attached to both ends of the system. The fixed boundary 
condition is assumed in the transverse direction. 
For the realization of 
the isotropic correlation for the magnetic fields 
in the sample region $L\times M$, we consider 
an additional regions of the width $5\eta_{\phi}$ 
outside of the sample in performing 
the summation over the uncorrelated fluxes $\delta \phi_i^{\rm u.c.}$.   
By this procedure, we realize isotropic correlations of disordered 
fluxes in the sample region\cite{KOOKSK}.

The two-terminal conductance $G$ is obtained by means of the 
Landauer formula
\begin{equation}
 G = (e^2/h) {\rm Tr }T^{\dagger}T,
\end{equation} 
where $T$ is the transmission matrix, which is evaluated by the transfer 
matrix method\cite{PMR}. 
The width $M$ is set to be 20. 
The smallest magnetic flux $\phi/\phi_0$ per plaquette
is $1/32$, and the corresponding magnetic length 
$l=\sqrt{\hbar/eB}=\sqrt{\phi_0/(2\pi \phi)}$ is
about $2.3$, much smaller than the system width $M$. 
This leads to the existence of edge states 
along the boundaries of the present systems.

\section{Two-terminal conductance}

\begin{figure}
\includegraphics[scale=0.7]{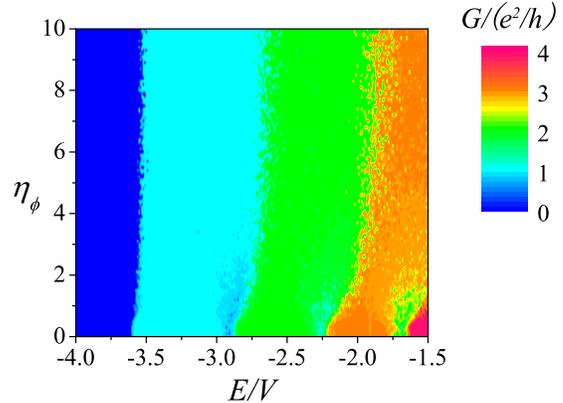}
\caption{(Color online)
Conductance for $w_{\phi}/\phi_0 =1/32$, $\phi/\phi_0 = 1/16$ 
and $L/M=50$ as a function 
of the correlation $\eta_{\phi}$ 
and the Fermi energy $E/V$. The magnetic length for the mean value 
$\phi$ is $l\approx 1.6$. Independent sample is assumed for each set of
parameters $E/v$ and $\eta_{\phi}$. 
\label{fig2}
}
\end{figure}

\begin{figure}
\includegraphics[scale=0.7]{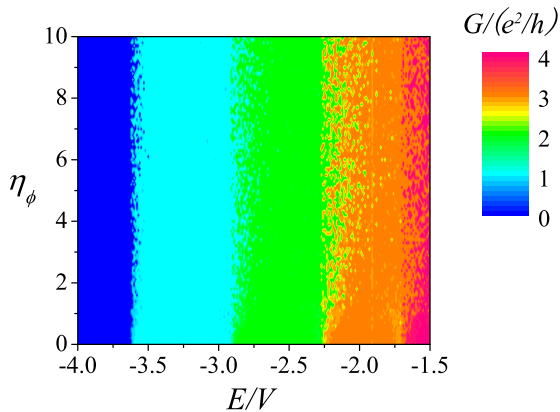}
\caption{(Color online)
Conductance for $w_{\phi}/\phi_0 =1/32$, $\phi/\phi_0 = 1/16$ 
and $L/M=2$ as a function 
of the correlation $\eta_{\phi}$ 
and the Fermi energy $E/V$. The distance between the plateau transitions
is insensitive to the strength of the correlation. Independent sample is 
assumed for each set of parameters $E/V$ and $\eta_{\phi}$. 
\label{fig3}
}
\end{figure}

\begin{figure}
\includegraphics[scale=0.7]{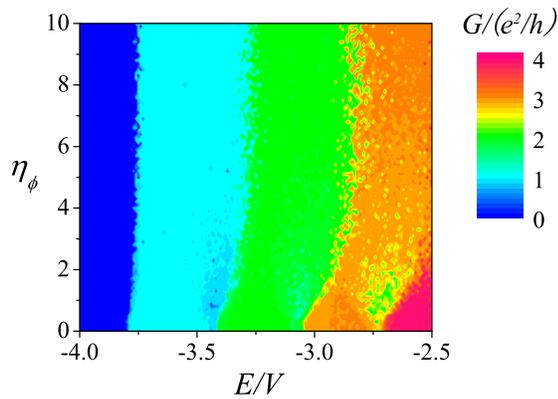}
\caption{(Color online)
Conductance for $\phi/\phi_0 = 1/32$, $w_{\phi}/\phi_0 =1/64$ 
and $L/M=50$ as a function 
of the correlation $\eta_{\phi}$ 
and the Fermi energy $E/V$. The distance between the plateau transitions
is increased in the presence of correlation. Note that the plotted 
energy range is different from Figs. \ref{fig2} and \ref{fig3}. 
The magnetic length for 
$\phi$ is $l\approx 2.3$. Independent sample is 
assumed for each set of parameters $E/V$ and $\eta_{\phi}$.
\label{fig4}
}
\end{figure}

We show in Fig. \ref{fig2} the conductance of a 
long wire ($L/M=50$) as a function of the Fermi energy $E/V$ and 
the correlation length $\eta_{\phi}$ of the random magnetic fields. 
The parameters for the magnetic fields 
are $\phi/\phi_0 =1/16$ and $w_{\phi}/\phi_0=1/32$. 
Independent samples are assumed for each set of parameters $\eta_{\phi}$
and $E/V$ throughout this section. It is 
clearly seen here that the plateau transitions shift toward higher energies 
as the correlation of 
random magnetic fields is increased. It should be noted 
in Fig. \ref{fig2} that the 
distance between successive transitions, namely the plateau width, 
increases as the correlation of 
magnetic fields is increased. 
This means that the shift of the critical energy of the plateau 
transition associated with 
lower Landau levels seems to be smaller than that associated with higher 
Landau levels.  
It is to be recalled that in the case of  
the correlated disorder potential\cite{KOOKSK}, 
the scale of shifts is common to all plateau transitions and 
the distance between plateau transitions remains to be the same as the bulk 
Landau level separation (Fig. \ref{fig1}).
The change of the critical energy in the random field case 
is therefore qualitatively different from that 
observed for the case of the potential correlation. 
This change in the separations of 
plateau transitions is absent for a short system (Fig. \ref{fig3}), and 
hence this change is a specific feature to long wires.
Insensitivity of the critical energy to the correlation in a short 
system is consistent with the fact that the positions of the 
bulk Landau levels are insensitive to the random magnetic field 
correlation as long as 
the disorder is rather weak \cite{Yakubo}.  
For a long wire, 
the present enhancement of the plateau width with increasing correlation 
is also observed for the case of 
$\phi/\phi_0 = 1/32$ and $w_{\phi}/\phi_0 = 1/64$
(Fig. \ref{fig4}). This is therefore a general feature of the 
conductance along long wires in spatially correlated magnetic 
fields. In both cases, the crossover is likely to take place 
when the correlation length $\eta_{\phi}$ is of the order of 
the magnetic length $l$.

It is also to be mentioned in Fig. \ref{fig2} 
that the CMIT for small correlation length $\eta_{\phi}$
does not necessarily occur at every bulk Landau level. In this case, 
the conductance does not vanish at higher Landau levels even in the 
absence of correlation (see also Fig. 5).  
This indicates that in the present case, 
the mixing of the edge states and the bulk states 
is not strong enough to reduce the conductance to be zero.
Note that in the case of 
the potential disorder,  
the conductance vanishes at the centers of 
the bulk Landau levels when the correlation is weak \cite{KKO,SKOK,KOOKSK}.  
In Fig. \ref{fig4}, it is also confirmed that 
the signature 
of CMIT survives for larger correlation length $\eta_{\phi}$ than 
in the case of Fig. \ref{fig2}, because  
the magnetic length is larger.

In order to clarify the origin of the present increase of 
the plateau width in the presence of the correlation of 
the random components of the magnetic fields, we investigate the 
positions of the critical energy more carefully.
We find that in the absence of correlation, 
the critical energies coincide with the
centers of the bulk Landau subbands. The plateau width is therefore 
determined by the mean value $\phi$ of the disordered magnetic fields.
On the other hand, in the presence of correlation, it is likely that 
the plateau width becomes larger and is determined by a stronger field 
$\phi + w_{\phi}/2$, which is effectively the maximum of the 
distribution of disordered magnetic fields.

\begin{figure}
\includegraphics[scale=0.6]{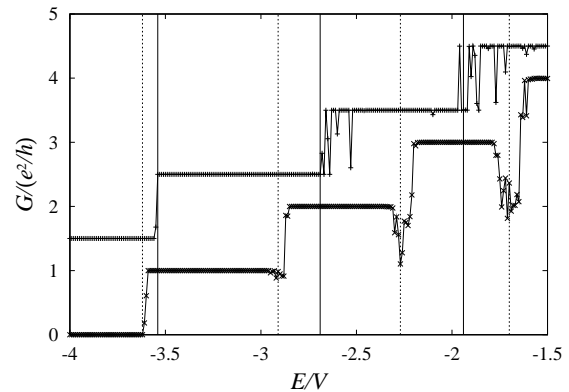}
\caption{Conductance for $\eta_{\phi} = 0(\times)$ and 
$\eta_{\phi}=5(+)$ 
in the case of $L/M=50$, $\phi/\phi_0 = 1/16$ and $w_{\phi}=1/32$ 
as a function 
of the Fermi energy $E/V$. For $\eta_{\phi}=5(+)$, $G/(e^2/h)+1.5$ is plotted 
instead of $G/(e^2/h)$. The vertical lines and dotted lines 
represent the positions of the critical energies in the absence of 
disorder ($w_{\phi}=0$) for $\phi/\phi_0 = 1/13$ 
and $\phi/\phi_0 = 1/16$, respectively. Independent sample is 
assumed for each value of $E/V$.
\label{fig5}
}
\end{figure}

In Fig. \ref{fig5}, 
the conductances for $\eta_{\phi}=5$ and for $\eta_{\phi}=0$ are 
shown.  The parameters are assumed to be the same as those 
in the case of Fig. \ref{fig2}, 
namely $L/M=50$, $\phi/\phi_0 =1/16$ and $w_{\phi}/\phi_0=1/32$.
Note that the correlation length $\eta_{\phi}=5$ is larger than 
the magnetic length for the mean value $\phi$ of the magnetic field.
The vertical lines represents the positions of the 
critical energies for $w_{\phi}=0$
and $\phi/\phi_0 = 1/13$. This value of the magnetic field is effectively 
the maximum value realized in the present system since 
$1/13 \approx (\phi + w_{\phi}/2)/\phi_0= 1/16 + 1/64 $. 
The vertical dotted lines are 
those for $w_{\phi}=0$ and $\phi/\phi_0 = 1/16$, equivalent to the
positions of the bulk Landau subbands. 
The results clearly suggest that the critical energy 
in the presence of correlation ($\eta_{\phi}=5$) is in good agreement with the 
Landau level position for the case of $w_{\phi}=0$
and $\phi/\phi_0 = 1/13$. It is to be emphasized that when the correlation is 
switched off ($\eta_{\phi}=0$), the critical energies  
coincide with the Landau level positions for the mean value of the fields 
$\phi/\phi_0 = 1/16$ instead of $\phi/\phi_0 = 1/13$, even in the case of 
$w_{\phi} \neq 0$. These facts mean 
that the conductance plateau transitions of a long wire in the 
presence of correlation is effectively determined by the stronger field 
$\phi + w_{\phi}/2$.

\begin{figure}
\includegraphics[scale=0.6]{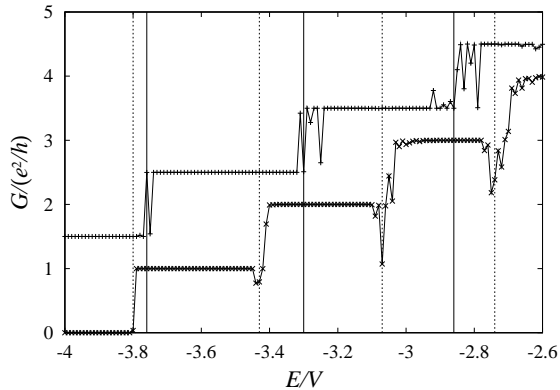}
\caption{Conductance for $\eta_{\phi} = 0(\times)$ and 
$\eta_{\phi}=5(+)$ 
in the case of $L/M=50$, $\phi/\phi_0 = 1/32$ and 
$w_{\phi}/\phi_0 =1/64$ as a function 
of the Fermi energy $E/V$. For $\eta=5(+)$, $G/(e^2/h)+1.5$ is plotted 
instead of $G/(e^2/h)$. The vertical lines and dotted lines 
represent the positions of the critical energies in the absence of 
disorder ($w_{\phi}=0$) for $\phi/\phi_0 = 1/32 + 1/128$ 
and $\phi/\phi_0 = 1/32$, respectively. Independent sample is 
assumed for each value of $E/V$.
\label{fig6}
}
\end{figure}

We have carried out the same analysis also for a different 
strength of the magnetic fields 
$\phi/\phi_0 = 1/32$ and $w_{\phi}/\phi_0 = 1/64$.
The results are shown in Fig. \ref{fig6}. Here we find again that the 
critical energies in the presence of correlation $\eta_{\phi}=5$
agree with the positions of the Landau levels in 
the effective magnetic field $\phi/\phi_0 = 1/32 + 1/128$.
On the other hand, for uncorrelated case ($\eta_{\phi}=0$), 
the critical energies 
are identical to the Landau level positions for the mean value 
$\phi/\phi_0=1/32$.
All these results support the conclusion that for the case of the correlated 
magnetic fields, the critical energies of 
the conductance plateau transition in long wires are determined by the 
effective field $\phi + w_{\phi}/2$, which is of the order of the 
maximum value of the disordered magnetic fields in the system. 
The plateau width therefore becomes larger, accordingly.

\section{Conductance of individual samples} 

\begin{figure}
\includegraphics[scale=0.6]{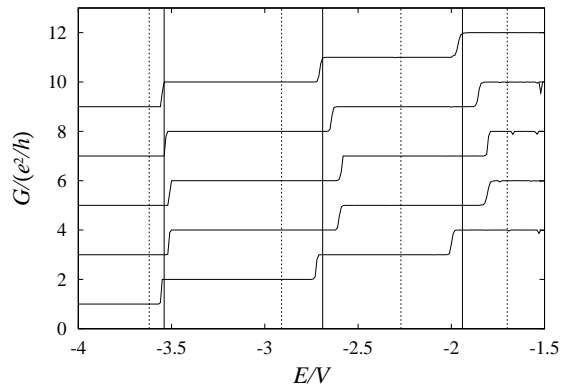}
\caption{Conductance for 5 samples 
shown with different offsets. The conductance in the region 
$E/V < -3.8$ is zero for all samples. The parameters are 
taken to be $\eta_{\phi}=5$, 
$L/M=50$, $\phi/\phi_0 = 1/16$ and 
$w_{\phi}/\phi_0 =1/32$. The vertical lines and dotted lines 
represent the positions of the critical energies in the absence of 
disorder ($w_{\phi}=0$) for $\phi/\phi_0 = 1/13$ 
and $\phi/\phi_0 = 1/16$, respectively.
\label{fig7}
}
\end{figure}

In Figs. \ref{fig5} and \ref{fig6}, 
the fluctuation of the conductance around the 
critical energies is seen, especially at higher energies.
To see the origin of this fluctuation of the conductance, 
we have calculated the conductance for 
various energies keeping the magnetic field configuration fixed.
The results for 5 samples of different random magnetic field configurations 
are shown in Fig. \ref{fig7}.
Here it is clearly seen that the fluctuation of the conductance seen 
in Fig. \ref{fig5} is due to the 
sample dependence of the critical energies, which originates from the 
plateau width dependence on the sample.
The critical energies therefore depend on the sample particularly at 
higher steps. This plateau width 
dependence is naturally understood as a consequence of the 
sample dependence of the effective field which determines the 
critical energies of the system. As we have shown in the previous section,
the effective field is of the order of the maximum value 
of the disordered magnetic field in the system. 
The critical energy $E_c$ would be determined by  
the effective field as  
\begin{equation}
 E_c = (n+1/2)\hbar\omega_{\rm eff}, 
\end{equation}
where $\omega_{\rm eff}$ denotes the cyclotron frequency corresponding to 
the effective magnetic field.
The fluctuation of the effective field is equivalent to the fluctuation of
$\omega_{\rm eff}$, which is independent of the energy. The fluctuation 
of $E_c$ is therefore larger for larger $n$. 
It is to be noted that this plateau width dependence on the sample 
has not been 
observed for the case of the correlated potential \cite{KOOKSK2}.

\begin{figure}
\includegraphics[scale=0.7]{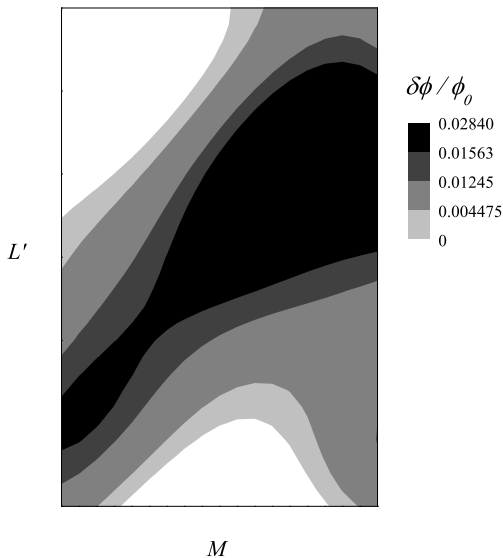}
\caption{The magnetic field landscape of a sample (the second 
from the bottom of Fig. \ref{fig7}) with 
$\phi/\phi_0=1/16$, $w_{\phi}/\phi_0 = 1/32$ and $\eta_{\phi}=5$. 
The part ($L'=30$), which is 
crucial to the conductance of the whole system, is shown. 
The horizontal (vertical) direction is across (along) the wire. The region 
where the magnetic field is stronger than its average is shown 
by the gray scale. In particular, the  
magnetic fluxes larger than $\phi + w_{\phi}/2$ are 
presented by the black region.
\label{fig8}
}
\end{figure}

\begin{figure}
\includegraphics[scale=0.6]{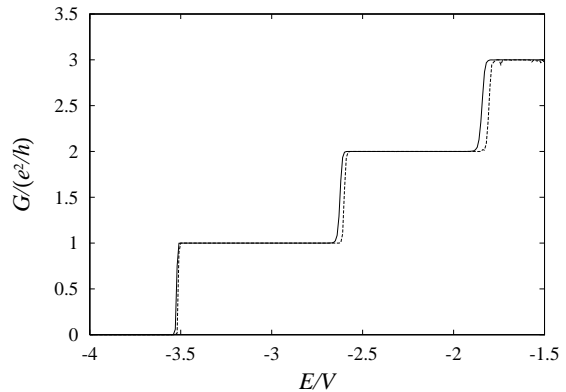}
\caption{Conductance of the small part presented 
in Fig. \ref{fig8} (solid curve). The conductance of the whole system 
is also presented by the dashed curve for comparison. 
The excellent coincidence between 
the solid and the dashed curves clearly proves that 
this small part 
effectively determines the conductance of the whole system.
\label{fig9}
}
\end{figure}

Now we discuss what determines the effective field of the sample which 
governs the conductance plateau transitions of a long wire in correlated 
disordered magnetic fields.
Let us consider the case of the second sample from the bottom in Fig. 
\ref{fig7}.
We show in Fig. \ref{fig8} the magnetic field landscape of 
the important part of the sample. 
The length $L'$ of this part is only $L'=30$.
Remarkably, this small part 
effectively determines the transport of the whole system ($L=1000$).
We evaluate the conductance of this small part separately and 
compare it with the 
conductance of the whole system (Fig. \ref{fig9}). 
In spite of the 
fact that the length of this small part is only 3 percent of 
the whole system, the 
conductance steps are almost reproduced. It is therefore clear 
that this small part is crucial to the conductance of the whole wire and 
that the effective field, which governs the conductance plateau transition,
is determined by the magnetic field strength of this part.

It is important to note that in this part, 
a region where the magnetic field is stronger  
than the mean lies across the wire (Fig. \ref{fig8}).
In such a region, the number of edge states is less than the average. 
It is therefore natural to expect that 
such bridge structure of a stronger magnetic field region 
acts as an effective barrier against the transport along
the wire and the conductance is determined by the number of edge states
there. If the correlation is absent, the typical 
size of the region with stronger magnetic field is 
of the order of the lattice constant and the probability to have such a 
region across the wire is exponentially small as has been discussed in 
our previous paper\cite{KOOKSK}. 
For the case of short systems, it is obvious that the probability is 
also small. In fact, this bridge structure of a stronger magnetic 
field region appears only once or twice 
in the sample with $L/M=50$ in the case of $\eta_{\phi}=5$. 
This argument naturally 
explains why the increase of the plateau width occurs
only in a long wire with a large correlation length $\eta_{\phi}$. 
The probability to have such a region across the wire is essentially 
the same as that for the potential barrier 
discussed in our previous paper \cite{KOOKSK}.

\section{Summary and concluding remarks}

We have investigated numerically the conductance
of the quantum Hall wires with correlated magnetic fields and examined 
how the effect of correlation shows up in the conductance plateau 
transition. It has been clearly demonstrated that in the presence of 
correlation $\eta_{\phi} > l$ 
the plateau width becomes larger than that for the 
uncorrelated case. It has been shown that the larger plateau width in 
the presence of correlation 
is determined by the 
effective field, which is of the order of the 
maximum value of the magnetic field in the system.
In particular, we have shown that the bridge structure of the high 
magnetic field region across the wire essentially 
determines the effective field and 
therefore governs
the conductance plateau transition of long quantum Hall wires. 
The change in the critical energy of the plateau transition is  
qualitatively different from that 
for the correlated potential case, where the plateau width 
is unchanged by the potential correlation \cite{KOOKSK}.
In spite of this apparent difference,  
the change of the plateau width can also be understood semi-classically 
by the fact that the number of edge states for a given 
energy at the bridge structure is smallest and therefore 
effectively determines the conductance of the whole system.

In the present paper, we have focused on the plateau transitions in 
the case of the correlated magnetic field, which 
is more relevant for actual experiments than the 
uncorrelated case. Before concluding this paper, we make a
remark about the CMIT in the case of the uncorrelated magnetic field.
In Fig. 10, the conductances in the uncorrelated magnetic field
are shown for the cases of $w_{\phi}/\phi_0 = 1/16$ and of 
$w_{\phi}/\phi_0 = 1/32$. The uniform component 
$\phi/\phi_0$ is $1/16$ in both cases. 
For the case of 
the larger fluctuation $w_{\phi}/\phi_0 = 1/16$, the precursor of 
the CMIT can be seen. It is therefore likely that the 
fluctuation larger than the mean is necessary to obtain the clear 
insulating region.
On the other hand, the clear 
conductance plateaus would vanish when the fluctuation
is larger than the mean.
We have therefore confined ourselves to the case where the
fluctuation $w_{\phi}$ of the magnetic field 
is smaller than its mean $\phi$ to discuss the plateau transitions.
It would be worth pointing out that 
in the previous numerical studies of the transport properties 
in the two-dimensional disordered magnetic fields  
\cite{SN,KO}, the scattering by 
the random magnetic fields has been shown to be weak and 
quantitatively different from that by the random potential.  

\begin{figure}
\includegraphics[scale=0.6]{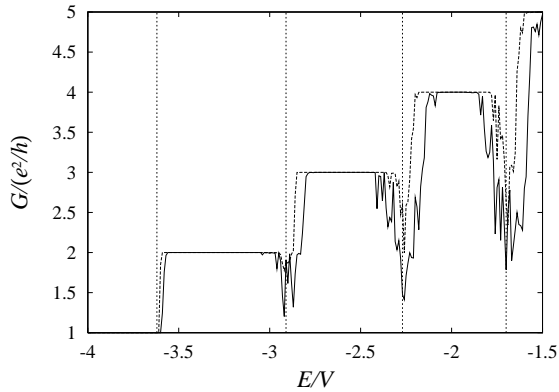}
\caption{Conductance of the quantum Hall wires with uncorrelated random 
magnetic fields $\eta = 0$ as a function of the Fermi energy $E/V$. 
The length of the system is $L/M=50$ and  
the uniform component of the magnetic field is $\phi/\phi_0 = 1/16$.
The solid curves and the dashed curves represent the cases for 
$w_{\phi}/\phi_0 = 1/16$ and for $w_{\phi}/\phi_0 = 1/32$, respectively.   
The vertical dotted lines 
indicate the bulk Landau level positions in the absence of 
disorder ($w_{\phi}=0$) for $\phi/\phi_0 = 1/16$. The precursor of the 
CMIT can be seen around these energies. 
\label{fig10}
}
\end{figure}

\begin{acknowledgments}
This work was partly supported by a Grant-in-Aid 
No. 16540294 and No. 18540382 
from the Japan Society for the Promotion of Science and by 
"Schwerpunktprogramm Quanten-Hall-Systeme", grant No. KE 807/2-1 
from the German Research Council (DFG). 
\end{acknowledgments}


\end{document}